\documentclass[twocolumn]{webofc}
\usepackage[varg]{txfonts}   
\wocname{EPJ Web of Conferences}
\woctitle{SuGAR2015--Searching for the Origin of the Galactic Cosmic Radiation}
\begin{document}
\title{Impact of Fermi-LAT and AMS-02 results on cosmic-ray astrophysics }
%
%

\author{Charles D.\ Dermer\inst{1}\fnsep\thanks{\email{charles.dermer@nrl.navy.mil}} 
}

\institute{Space Science Division, US Naval Research Laboratory, Washington, DC USA 
          }

\abstract{%
This article reviews a few topics relevant 
to Galactic cosmic-ray astrophysics, focusing on
the recent AMS-02 data release and
 Fermi Large Area Telescope data on the diffuse
Galactic $\gamma$-ray emissivity. Calculations are made of the diffuse 
cosmic-ray induced $p+p \rightarrow\pi^0\rightarrow 2\gamma$ spectra,
normalized to the AMS-02 cosmic-ray proton spectrum at $\approx 10 - 100$ GV, 
with and without a hardening in the cosmic-ray proton spectrum at 
rigidities $R \gtrsim 300$ GV. 
A single power-law momentum ``shock" spectrum  for the local 
interstellar medium cosmic-ray proton spectrum cannot be ruled out 
from the $\gamma$-ray emissivity
data alone without considering the additional contribution of electron 
bremsstrahlung. 
Metallicity corrections
are discussed, and a maximal range of nuclear enhancement factors from 
1.52 to 1.92 is estimated.
Origins of the 300 GV cosmic-ray proton and $\alpha$-particle hardening are discussed.
}
\maketitle
\section{Introduction}
\label{intro}

On the way to writing this proceedings article for  
the 2$^{nd}$ SUGAR (Searching for the sources of galactic cosmic rays)
Workshop, held in the D\'epartement de Physique nucl\'eaire et corpusculaire  
at  Universit\'e Gen\`eve on 21 -- 23 January 2015, the Alpha Magnetic 
Spectrometer Collaboration held a significant data release
on its  ``AMS days at CERN,"  which took place 15 -- 17 April 2015 \cite{AMS02}. 
Since new data is bound to make old theory more interesting,
I've redirected my thoughts towards this latest high-quality cosmic-ray 
data.

The major results
presented at the AMS days were the $\bar p/p$ ratio from kinetic energies $T_p \lesssim 1$ GeV to 
$T_p \approx 350$ GeV, and the cosmic-ray proton and $\alpha$ particle
spectra for rigidities 10 GV $\lesssim R\lesssim 1500$ GV (p) 
and 10 GV $\lesssim R\lesssim 2500$ GV ($\alpha$). Combining these
results with
the previous AMS-02 results on the positron fraction  \cite{2013PhRvL.110n1102A} and 
the cosmic-ray electron and positron fluxes \cite{2014PhRvL.113l1102A}, 
in addition of course to the vast wealth of cosmic-ray data \cite{Strong},
 gives
an unprecedentedly large and accurate data set for studying cosmic-ray physics.

In the most recent AMS-02 data \cite{AMS02}, the $\bar p/p$ ratio is reported to be roughly 
constant for $10$ GeV $\lesssim T_p$ $\lesssim 350$ GeV with a value
of $\approx 2\times 10^{-4}$. 
The measured $\bar p/p$ ratio is 
claimed to be a factor of $2$ -- $3\times$ larger than predicted
by secondary cosmic-ray production models constrained by, for example, 
the  positron fraction 
e$^+$/(e$^+$+e$^-$) and the B/C ratio \cite{2009PhRvL.102g1301D}.
Given the associated rising positron fraction 
 \cite{2013PhRvL.110n1102A} at $\gtrsim 10$ GeV,
which is often explained as due to additional sources like pulsars, and not from
enhanced cosmic-ray induced production, one may ask whether there
is a consistent explanation for the overabundant $\bar p$ 
and the enhanced positron fraction.

AMS-02 confirms, though at a higher value of 
break rigidity, the hardening in the cosmic-ray proton 
and $\alpha$ spectrum \cite{AMS02BoverC} reported 
from PAMELA results \cite{PAMELA}. The hardening appears in the
AMS-02 p and $\alpha$ spectra at $\approx 300$ -- 400 GV, 
compared to $\approx 240$ GV for cosmic-ray p and $\alpha$ 
spectra in PAMELA data \cite{2014PhR...544..323A}.
Nevertheless, as was remarked more than once at this workshop,
 there must be a hardening between $\approx 100$ GeV/nuc and the knee,
and it is great that AMS-02 has provided more precise details
(which improves upon the preliminary spectra released during
the 2013 ICRC \cite{AMS02}).

Rather than repeat the points made in my presentation, which 
is available at the conference 
website\footnote{https://indico.cern.ch/event/332221/session/5/contribution/38}
and reviews some of the work found in my previous conference papers 
\cite{2013arXiv1303.6482D,2013arXiv1307.0497D}, it seems 
more useful to examine a few of the implications of the AMS-02 data in 
view of the goals of this meeting. 

First we look at the  cosmic-ray proton and $\alpha$-particle spectra
from low energies to high energies, which represents the minimum local
interstellar medium cosmic-ray proton ($\alpha$)  (LISMCRp ($\alpha$)) spectrum
from which to calculate a minimum $\gamma$-ray emissivity. 

Second, we make some calculations of galactic emissivity from the cosmic-ray proton
data, using a simple nuclear enhancement factor $k_{nuc}$ to correct for metals.
Even with $k_{nuc} = 2$, a power-law momentum spectrum is allowed, so 
a contribution of bremsstrahlung $\gamma$ rays is required to infer
a low-energy break in  the
LISMCRp spectrum from the Fermi-LAT $\gamma$-ray data. A brief
discussion of metallicity ($Z>2, A>4$) corrections, including cosmic-ray and ISM metals,
is given.
We consider whether the breaks in the spectra of cosmic-ray protons and ions
are a consequence of propagation effects, 
multiple injection sources with different spectral indices, 
or due to hardening produced by nuclear collisions which preferentially depletes
the low-energy particles that traverse larger grammages.



\section{Cosmic rays and $\gamma$ rays}
\label{sec-1}

\subsection{Cosmic-ray p and $\alpha$ spectra}
\label{sec-1-2}

\begin{figure}
\centering
\includegraphics[width=7.0cm,clip]{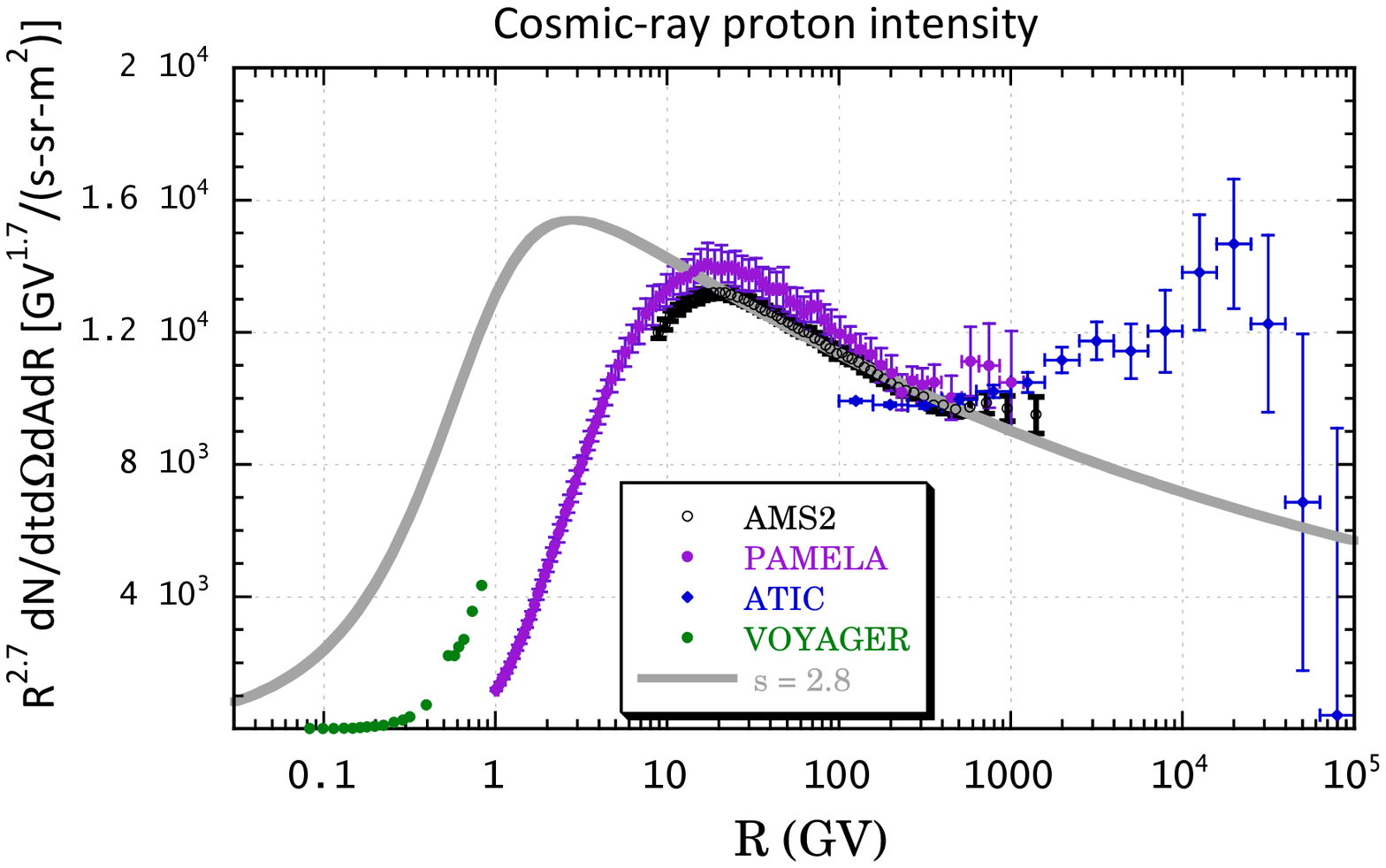}
\includegraphics[width=7.0cm,clip]{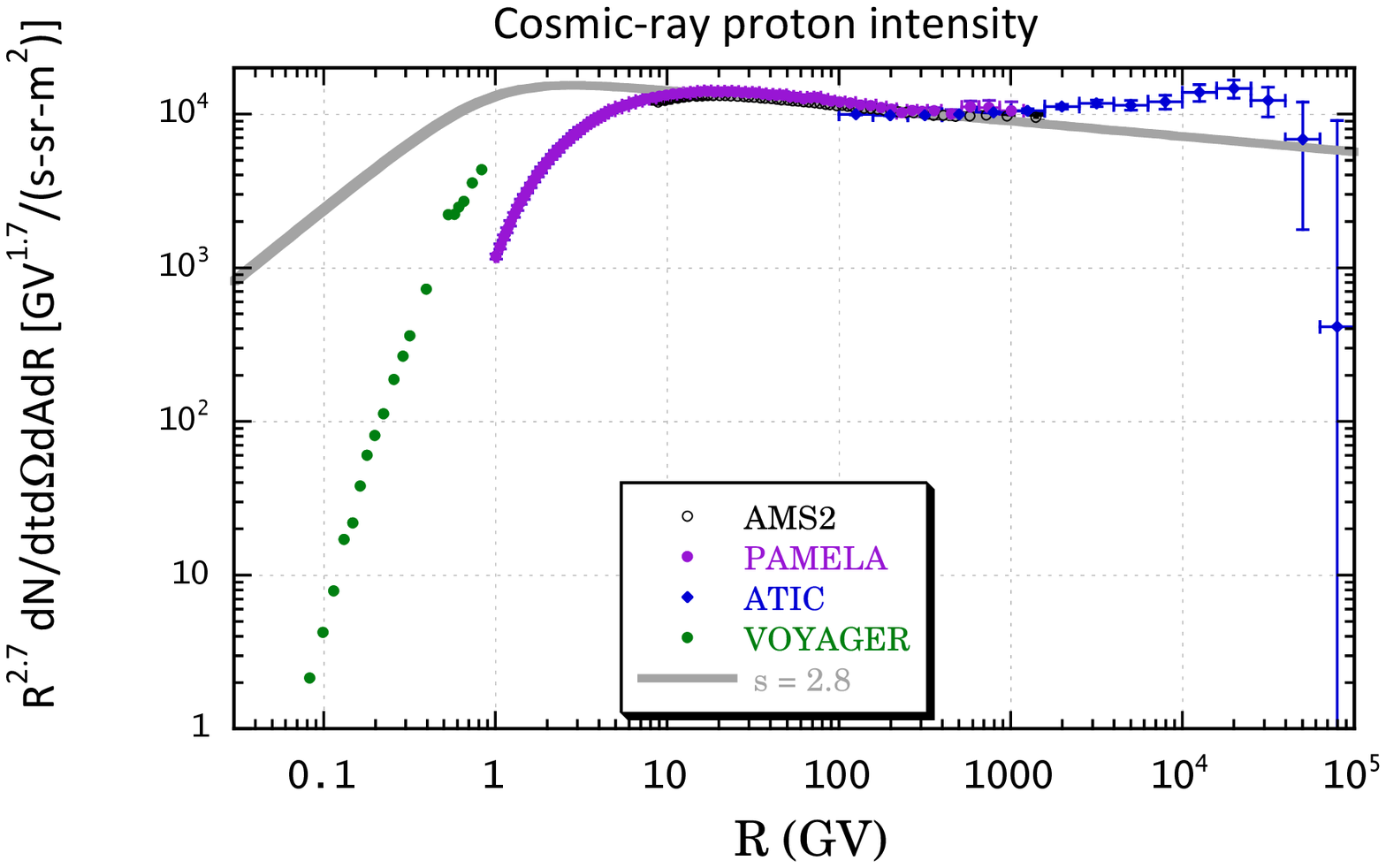}
\caption{Data, as labeled,  are from ATIC  \cite{Pa06}, 
PAMELA  \cite{PAMELA}, AMS-02 \cite{AMS02}, and
Voyager 1  \cite{Voyager1}, expressed in the form
$R^{2.7}\times$ particle number intensity. Smooth curves show
a power-law proton flux in momentum with index
$s=2.8$ in the same representation. Upper and lower panels show the 
fluxes on a linear and logarithmic scale, 
respectively. 
}
\label{fig-1}       
\end{figure}

In Fig.\ \ref{fig-1}, we construct a plot of the cosmic-ray proton intensity (or flux)
from 4 sets of data (see \cite{Strong}) measured at or near Earth and at $\sim 122$ AU (Voyager 1). 
The highest energy  data are from the ATIC (Advanced Thin Ionization Calorimeter)-2
balloon campaign during 2002 -- 2003 \cite{Pa06}. Also shown are  
PAMELA (Payload for AntiMatter Exploration and Light-nuclei Astrophysics) 
data \cite{PAMELA}, the recently reported AMS-02 data \cite{AMS02}, and
Voyager 1 data \cite{Voyager1}. One thing about which we can be quite certain
 is that due to Solar modulation of the cosmic ray,
the LISMCRp spectrum will be above the level of the data. Though 
it could be that the Voyager 1 has sampled the ``true" LISMCRp spectrum,
it is also possible that during the billions of years that the Solar wind 
has been generated, a much more complicated and extended magnetic
structure around the Sun has formed. In any case, the data should provide
a solid lower limit to the LISMCRp spectrum. Another thing is that 
AMS-02 and PAMELA differ in normalization by $\lesssim 10$\%, and more like
$\sim 5$\% at $10 \lesssim $R(GV)$\lesssim 100$, where modulation effects
are believed to be small. These differences are less than uncertainties
in nuclear cross sections \cite{2012PhRvD..86d3004K}.

Fig.\ \ref{fig-1} is a plot of the LISMCRp flux multiplied by $R^{2.7}$,
where $R\equiv P/Z$ is the particle rigidity, $P = \sqrt{E^2-m^2}$ is its
momentum, $E$ is its total energy, and $T = E-m$ is the particle kinetic energy.
For protons, then,
\begin{equation}
R^{2.7}\,{dN\over dAdtd\Omega dR} = {R^{3.7}\over E T^2_p(R)}\,\left[ T_p^2 \,{dN\over dAdtd\Omega dT_p}\right]\;.
\label{R27dN}
\end{equation}
The upper and lower panels show the measured cosmic-ray proton flux on a linear 
and logarithmic scale, respectively. There are several points to 
be made. First is that the flux, as shown in this representation multiplied by 
$R^{2.7}$, deviates barely by a factor of 2 and is 
remarkably close to a power law from $10\lesssim R$ (GV) $ \lesssim 10^4$.
Second is that three experiments---ATIC, AMS-02, and PAMELA---now converge on a break rigidity for 
protons of $R \approx 300$ GV, with the low-energy ($10\lesssim R$ (GV) $ \lesssim 300$) 
slope equal to $\approx 2.8$, and the high-energy ($300\lesssim R$ (GV) $ \lesssim 10^4$)
slope  $\approx 2.6$, based primarily on the ATIC data (the high-energy extension of 
the AMS-02 data suggests that the hardening is less, perhaps to a slope of $\approx 2.7$).

\subsection{The cosmic-ray spectrum in the local interstellar medium}
\label{sec-1-2}

What, then, is the LISMCRp spectrum and how is it made? 
The prevailing theory is that Galactic cosmic
rays are accelerated by processes taking place in SNR
shocks as the expanding supernova ejecta overtakes and sweeps up 
material in the surrounding medium \cite{2013A&ARv..21...70B}. 
This theory has been taught so much 
at school that to even question the SNR origin of Galactic 
Cosmic Rays at this stage is treated as heresy.\footnote{Rather than burnt at the 
stake, one's proposal isn't funded.} In any event,
this theory 
has received strong support from 
Fermi-LAT and AGILE observations of the $\pi^0\rightarrow 2\gamma$ 
feature in IC 443 and W44 \cite{2013Sci...339..807A}, as well as in the spectrum 
of a 3rd Fermi-LAT SNR, W51C, reported at 
this workshop by Dr.\ Jogler. 

Although the $\pi^0\rightarrow 2\gamma$ decay hardening in the low-energy $\gamma$-ray
spectrum below several hundred MeV has now been reported,
confirming innumerable predictions and calculations going back to 
Ginzburg \& Syravotskii and Hayakawa,
 some open
questions remain.
For example, the SNR $\gamma$-ray spectral indices 
can be soft (i.e.,  steeper than $2.5$), and unusually
low high-energy cutoffs in the $\gamma$-ray spectra are inferred from 
Fermi-LAT and ground-based $\gamma$-ray detectors. For example, 
W51C and W44 have cutoffs below 10 GeV, whereas Cas A has a break near 1 TeV
and RX J1713.7-2942 has a break at $\approx 10$ TeV. 
If the $\gamma$-ray emission is due
to cosmic rays accelerated to the knee 
($\approx 3$ PeV), $\gamma$ rays at a factor 
of 10 less energy ($\approx 300$ TeV), are expected,
and no photons near that high are detected. (The highest
energy photons, at $\approx 100$ TeV, are observed in the steeply declining
part of the RX J1713.7-3946 SNR spectrum \cite{2007A&A...464..235A}.)

Putting these concerns to the side for the moment, we marvel instead that
the cosmic-ray proton spectrum is so close to a power law, 
as expected in first-order test-particle Fermi acceration
theory, as reviewed in \cite{be87}. One of the central results of this theory is that the distribution 
function of transmitted particles $f(p)\propto p^{-3r/(r-1)}$, 
so that $dN/dp \propto p^{-A_{tp}}$, where the test-particle
spectral index 
\begin{equation}
A_{tp}= {2+r\over r-1}\;,
\label{Atp}
\end{equation}
Here $r$ is the compression ratio which, for strong shocks in a medium
with adiabatic index of $5/3$, appropriate to a nonrelativistic monoatomic gas,
cannot exceed a value of $r =4$, implying a test-particle index of $A_{tp}\cong 2$
for a strong nonrelativistic shock. 

If the injection spectrum of cosmic-ray protons or ions is a power-law in momentum 
or rigidity, then what is the form of the intensity $j_{sh}(T_p ) = dN/dAdtd\Omega dT_p$
that enters into the calculations of $\gamma$-ray emissivity? Most simply, 
rigidity-dependent escape steepens the injection spectrum by the index $\delta$, 
so that the steady-state number spectrum $dN/dp \propto p^{-A_{tp} -\delta} \propto p^{-s}$,
where $s = A_{tp} + \delta$. The intensity
\begin{equation} 
j_{sh}(T_p )\propto \beta(T_p) \,{dN\over dT_p} 
\propto \beta(T_p)\,|{dp\over dT_p}|\left( {dN\over dp}\right) \propto p^{-s}\;
\label{jshTp}
\end{equation}
\cite{2012PhRvL.109i1101D}. In Fig.\ \ref{fig-1}, a power-law momentum spectrum, normalized
to the AMS-02 data at $10$ GV $\lesssim R\lesssim 200$ GV with index 
$s = 2.8$ is shown. It clearly disagrees with the Voyager 1 data, though as noted, this 
might not be a fatal concern if the Voyager 1 data does not yet sample the LISMCR spectrum.

In any case, however, one would have to be extremely naive to assume that 
a power-law momentum spectrum should describe the LISMCRp intensity. Besides
diffusive escape from the target-rich Galactic disk, 
convective Galactic winds from the superpositions of the hot
gases of O/B stars and supernovae will cause a systematic 
expulsion of cosmic rays to the halo of the Galaxy. 
MHD turbulence in the Galactic magnetic field
could cause systematic acceleration of cosmic rays during transport  \cite{smp07}.
Although breaks may be expected, demonstrating such a break
in the cosmic-ray proton spectrum has proven to be difficult, for reasons
of (1) limitations on our knowledge of the secondary pion distributions 
in p-p collisions, and (2) metallicity corrections on secondary pion 
production \cite{2012PhRvL.109i1101D,2012PhRvL.108e1105N}.

A crucial issue in the scenario that 
strong SNR shocks accelerate the Galactic cosmic rays should be noted.
Compression ratios $r \rightarrow 4$ imply injection indices of cosmic rays 
near $2$. In theory the bulk of the cosmic-ray population is swept downstream and is confined
to the remnant until late in the SNR's radiative evolution when the shock is not
so strong, at which time the cosmic rays
diffusively escape. The ``real" injection index of cosmic rays into interstellar 
space may be somewhat softer, $\approx 2.1$ -- $2.2$. On the other hand, the 
highest energy particles will have escaped near the Sedov time.
For these injection indices, the rigidity-dependent steepening would have to be
by $\delta \approx 0.6$ -- 0.7. This can be compared with analyses finding
$\delta \approx 0.44\pm 0.03$ and a source injection index of $2.34$
\cite{2015arXiv150403134G}
using the form $D(R)=D_0\beta (R/R_0)^\delta$ for the diffusion coefficient.
PAMELA analysis finds $\delta \cong 0.4$ \cite{2014ApJ...791...93A}.

The secondary to primary ratios of cosmic-ray nuclei
 can also be modeled by an average
escape grammage $\Lambda(R)$ through which  cosmic-ray particles with
rigidity $R$ pass  \cite{smp07}.
Analysis of cosmic-ray ion composition
often uses 
the empirical function
$\Lambda(R) = \beta \Lambda_0 (R/R_0)^{-a}$ for $R>R_0$ and 
$\Lambda(R) = \beta_0 \Lambda_0 $ for $R<R_0$, 
where $\beta = p/E = ZR/\sqrt{Z^2R^2 - m^2}$.
When fit to the B/C ratio, a typical parameter set fitting the data 
has $\Lambda_0 = 11.8$ g/cm$^{-2}$ and  $a = 0.54$, 
and $R_0 = 4.9$ GV/c, with a rigidity index of 
$2.35$. The value of $a$ essentially coincides with the value of $\delta$ in diffusion 
theory.

The value $\delta \approx 0.4$ is near the theoretically favored
value of $\delta \approx 1/3$ for Kolmogorov turbulence, but
observationally then requires a soft injection index $\approx 2.3$ -- 2.4.
In any case, cosmic rays must sample a wide range of environments, and it
may be surprising that the cosmic-ray proton spectrum is as smooth as it is.
Now we consider $\gamma$-ray constraints on 
deviations from a single momentum power law at the low- and high-energy 
parts of the Galactic cosmic-ray proton distribution.

\subsection{$\gamma$-ray constraints on cosmic rays}

We can ask if the power-law momentum spectrum shown in Fig.\ 
\ref{fig-1} can be ruled out by the $\gamma$-ray data.
The expression $R^{2.7}\,dN/dtd\Omega dA dR = 1.8\times 10^4 (R^{0.9}/E)$
GV$^{1.7}$/(s-sr-m$^2$) translates into a cosmic-ray proton flux
\begin{equation}
j_{sh}(T_p) = 1.8 R^{-2.8}\;({\rm s~sr~cm}^2 {\rm~GV})^{-1}\;.\;
\label{jshTp}
\end{equation}
(In comparison,  I \cite{1986A&A...157..223D}  used the expression 
$j_{sh}(T_p) = 2.2 E_p^{-2.75}\;({\rm s~sr~cm}^2 {\rm~GeV})^{-1}$,
with $E_p$ the total energy in GeV,
based on the cosmic-ray proton spectrum 
shown in Simpson's review \cite{1983ARNPS..33..323S}.)

It is a simple matter to plug the proton flux given by Eq.\ (\ref{jshTp})
into a code that calculates the $\gamma$-ray emissivity per H atom due
to pion-producing reactions of relativistic protons with protons at rest, and compare with 
the emissivity measured with the Fermi-LAT telescope. 
For this purpose, I use my original code \cite{1986A&A...157..223D}, including the 
kludge
between 3 and 7 GeV, which may be the source of some structure in the 
$\gamma$-ray production spectra. The improvements suggested  \cite{2013arXiv1307.0497D}
to properly characterize
the different isobars according to their exclusive cross sections, 
which would mitigate the effects of the kludge,
have not yet been fully implemented. 
We also make calculations using Kamae's model \cite{kam06}.
Notably, a nuclear enhancement
factor $k_{nuc} =  2.0$ is assumed.

\begin{figure}
\centering
\includegraphics[width=8.0cm,clip]{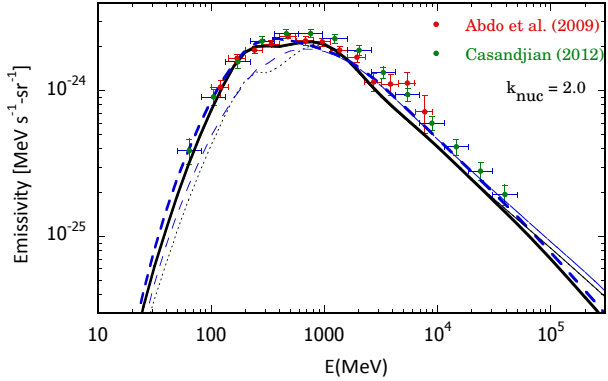}
\caption{Fermi-LAT data shows early results
of the Galactic $\gamma$-ray emissivity \cite{2009ApJ...703.1249A}, 
and the emissivity from the 2012 analysis of Casandjian  \cite{cas12}. Calculations
are for a proton momentum power-law shock spectrum given by eq.\ (\ref{jshTp}) (thick), 
the same spectrum with a hardening by $0.2$ units at 300 GeV (thin solid), 
and the shock spectrum with a low-energy cutoff at 2 GeV in the LISMCRp spectrum 
(thin dotted and dot-dashed); see Fig.\ \ref{fig-1}. 
Two $p+p\rightarrow \pi^0 \rightarrow 2\gamma$ models are used, shown in 
black  \cite{1986A&A...157..223D} and blue \cite{kam06}.
 }
\label{fig-2}       
\end{figure}

Fig.\ \ref{fig-2} shows calculations compared to Fermi-LAT data. The early 
Fermi-LAT emissivity study  \cite{2009ApJ...703.1249A} uses 
data from 2008 August 4 to 2009 January 31 ($\approx 6$ mo), 
whereas the 2012 study \cite{cas12} uses 3 years of Fermi-LAT data.
Besides the single momentum power-law shock spectrum, eq.\ (\ref{jshTp}),
we also  consider a case where
the shock spectrum hardens at $R_{br}= 300$ GV  by 0.2 units, 
and a case where the spectrum nose dives below 2 GeV.  
Even with Pass 8 analysis and 10 years of Fermi-LAT data,
it will be a big challenge to see a statistically significant
hardening above $\approx 30$ GeV. The two emissivities---with
and without the spectral hardening at $R > 300$ GV in the
LISMCRp spectrum---differ by only a factor of 2 at 100 GeV
photon energies. Attempts to see the spectral hardening from 
Earth-limb analysis may be more successful, but limitations of
our knowledge of secondary production cross sections can ultimately
hinder analysis.

This simple comparison does not preclude the 
possibility that eq.\ (\ref{jshTp}) could describe
the LISMCRp spectrum, at least below 300 GV. In fact, the 
numerical fit using the simple shock power-law momentum spectrum,
though going right through the low energy data, must allow
some emission for bremsstrahlung from nonthermal electrons making 
radio synchrotron and for bremsstrahlung from the  e$^+$ and e$^-$
made from the decay of $\pi^\pm$. A full study must consider associated
radio synchrotron emission \cite{2013MNRAS.436.2127O}, demodulation or reconstruction of the 
LISMCR e$^+$/e$^-$ spectrum, and the inverse Compton emissions 
that are dependent on Galactic location \cite{cas15,str15}.
A recent evaluation of $\gamma$-ray emissivity from the 
AMS-02 cosmic-ray proton and $\alpha$ data, including calculations of the associated neutrino
fluxes, see \cite{2015arXiv150400227G}.
Even so, emissions from leptons, because of their strong 
cooling, are expected to have softer spectra and contribute relatively less
to the $\gamma$-ray emissivity at higher
energies. It is not clear if the emissivity data is systematically 
$\sim 10$ -- 30\% larger than predicted by cosmic-ray induced 
emissions when using $k_{nuc} = 2$, and whether this is a 
reasonable value. This brings us to 
a reconsideration of the nuclear enhancement factor.

\subsection{Metallicity corrections}

The calculations shown in Fig.\ \ref{fig-2} make use of a simple multiplier
for the effects of all nuclei other than  protons and H. This is the 
nuclear enhancement factor $k_{nuc}$, which was assigned a value of 2.0 in our calculations.
Let us make a quick calculation of the range of values
of $k_{nuc}$.
The first thing to note is that in collision physics, $T_{nuc}$, 
the kinetic energy per nucleon is the quantity of interest. At the same
values of $T_{nuc}$ we  see that the accelerated particle distributions---the
cosmic rays---have abundances of H, He, CNO and Fe in the ratio 
of 1:0.07:0.0044:0.0003 \cite{2012PhRvL.109i1101D,smp07}. 
From Mori \cite{2009APh....31..341M}, drawing on the work of  J.-P.\ Meyer \cite{Meyer},
we take the composition of the target ISM material
with relative abundances of H:He:CNO:NeMgSiS:Fe in the ratio
1:0.096:1.38e-3:2.11e-4:3.25e-5.

\begin{table}
\centering
\caption{Relative contributions to nuclear enhancement factor $k_{nuc}$}
\label{tab-1}       
\begin{tabular}{lll}
\hline
~ & $\sigma\propto A_1A_2$ & $\sigma\propto A^{2/3}_1A_2^{2/3}$  \\\hline
p-H & 1 & 1 \\
p-He + $\alpha$-H & 0.66 & 0.42 \\
$\alpha$-He   & 0.11 & 0.043\\
CNO & 0.11 & 0.042\\
NeMgSiS  & 0.014 & 0.0045\\
Fe & 0.022 & 0.0042\\
Total& 1.92 & 1.52\\\hline
\end{tabular}
\end{table}


The maximum enhancement due to nuclei can be estimated assuming that there is no shadowing, 
and each nucleon participates independently in the scattering. On the other hand, 
an apparent minimum value of $k_{nuc}$ can be obtained by assuming a geometrical 
shadowing $\propto A^{2/3}$, as if the nucleons were tightly packed in the nucleus. 

Table~\ref{tab-1} shows the relative contributions of various nuclei or nucleon groups
to $k_{nuc}$, under the two extremes that $\sigma\propto A_1A_2$ and $\sigma \propto 
A_1^{2/3}A_2^{2/3}$. 
The notation is such that the row labeled CNO includes all reactions of 
C, N, and O in the cosmic rays and in the ISM involving p, He, and $\alpha$ particles.
Similarly, NeMgSiS includes all reactions of CR and ISM Ne, Mg, Si, and S in the cosmic
rays and ISM that interact with p, $\alpha$, and CNO. For this estimate, we assume the 
same cosmic ray as ISM composition for NeMgSiS. 

Note that reactions involving ions heavier that He account for up to 5\% of the emissivity,
so cannot be in any sense neglected. 
This estimate shows that $k_{nuc}=2.0$ exceeds the upper limit, so 
that the emissivity calculations in Fig.\ \ref{fig-2} are, if 
anything, an overestimation.
Mori \cite{2009APh....31..341M}, using DPMJET-3 to derive cosmic-ray energy-dependent emissivities,
does not give results below 6 GeV/nuc. 
Recent treatments are given by \cite{2014ApJ...789..136K} who, using different
compositions, find $1.9 \lesssim k_{nuc} \lesssim 2.1$, in better agreement
with the ``no-shadowing" cross sections. As noted there, the situation is more complicated
than a simple value of $k_{nuc}$, which is dependent on the photon energy and the 
differences between cosmic-ray proton and ion spectra.
Other treatments find corrections $\propto A^{2.2/3}$ 
\cite{2007NIMPB.254..187N}. The low value  of $k_{nuc} = 1.45$ 
obtained in my early study \cite{1986A&A...157..223D}
was a consequence of using a cross-section correction closer to the $A^{2/3}$
behavior, and  by using a low value for the  cosmic-ray $\alpha$ particle to proton ratio.

\subsection{Cosmic-ray spectral hardenings}

Cosmic-ray spectral 
hardenings at $\approx 300$ GV in 
the p and $\alpha$ spectra could be 
due to any number of physical effects. 
From general principles, we can suggest
either that this spectral 
hardening is due to (1) the superposition 
of two populations with different
mean injection indices; (2)
rigidity-dependent effects on escape from the 
Galaxy; 
(3) rigidity-dependent effects at the source/accelerator;
(4) acceleration during transport; 
or (5) grammage effects.

Effect (5) fails because grammage makes instead a low-energy
hardening when heavier ($\gtrsim$CNO) low-energy 
cosmic-rays pass through significant
grammages during escape. Effect (4), which 
suggests turbulent reacceleration, would 
 make curved, not broken power-law spectra \cite{2006ApJ...647..539B}.
Effects (2) and (3) go hand-in-hand, in terms of the behavior
if not in explanation. All Fermi acceleration theories
have, in the collisionless  limit, a dependence
on rigidity as the fundamental quantity. A break
or hardening at a given rigidity should manifest 
as one of the 3 (Bernard) Peters cycle predicting $E_{br}\propto
Z$, as described 
by Dr.\ Tilav at this workshop. 

Idea (1) seems to be ruled out from a number of directions.
A 0.2 break from the addition of two underlying distributions
that only differ by 0.2 in index would be extremely broad, 
and there would be unusual changes in composition depending
on the compositions of the 2 populations. Which returns us to (2) and 
(3), for which we need an explanation why, during cascading
from small to large $k$, 
the wave turbulence spectrum governing diffusion would 
apparently soften, so that lower energy ions that gyroresonate
with these waves would escape more readily than at high energies
resulting in a hardening.


\subsection{$\bar p/p$ ratio}

The figure in the AMS-02 release \cite{AMS02} 
showing a severe disagreement  of the 
$\bar p/p$ ratio  with a 
secondary production model 
prediction deserves a few comments.
First is that ratios $\bar p/p$ and 
$e^+/(e^++e^-)$ are far less useful
than the particle spectral intensities themselves \cite{2014JCAP...09..051K}, 
particularly when looking at fits to data
using models
involving multiple source
 populations. Second is that the model
\cite{2009PhRvL.102g1301D} showing
a discrepancy with the AMS-02
data, used to constrain
fits for dark matter candidates to 
$\bar p$ production, was 
tailored to  fit PAMELA
data that becomes increasingly uncertain
at high energies. 

If there are $\bar p/p$ enhancements compared to 
expectations from secondary production,
a non-Copernican explanation has us fortuitously
situated some 2 Myr ago next to a SN that deposited
$^{60}$Fe while enhancing cosmic-ray production in 
$\bar p$ and e$^+$
 \cite{2015arXiv150204158K}.
Fits to the $\bar p/p$ data with a conventional
propagation parameters 
($1/3 < \delta\lesssim 1/2$)
in a 
secondary nuclear production models
are found in recent analyses \cite{2015arXiv150404276G,2015arXiv150407230L}. 
In particular, agreement with the B/C and $\bar p/p$ 
ratio with a propagation model with $\delta \approx 0.42$ is
found in \cite{2015arXiv150405175E}.

\subsection{Secondary nuclear production models}

The high-quality Fermi-LAT data demands increasingly
accurate nuclear data, which is currently lacking. 
Some new low-energy, $\lesssim 1$ GeV/nuc data are 
presented in \cite{2014PhRvD..90l3014K} and were 
discussed at this workshop by Dr.\ Taylor.  
Another paper that would be a valuable benchmark
for low-energy secondary production calculations 
is \cite{1988JPhG...14..903A}. All the new 
data is at $T_{nuc} \lesssim 2$ GeV/nuc. 
The important regime
for $\gamma$-ray astronomy is at 
$\approx 2$ -- 10 GeV/nuc, where scaling models
break down, event generators extrapolate
outside their zone of certainty, and 
exclusive formulations become impossible.

Rather than providing empirical formulations
that are specific, model-dependent, and can't
 be tampered with, a better direction
is to have physical models that can be modified.
Detailed examination of processes in the regime
is required. 
Intermediate-energy nuclear astrophysics is
not, unfortunately, a growth area.

A relevant formula to conclude this 
contribution 
is the $\pi^0$-production 
threshold kinetic energy per nucleon $T_{thr,nuc}$
of a cosmic-ray ion with mass $M = A_1m_p$ 
striking an interstellar atom or ion with mass $m = A_0m_p$.
The threshold condition is that $s = E^2 - p^2  = 
(m+M+m_{\pi^0})^2$. I obtain
\begin{equation}
T_{thr,nuc}= {m_\pi}\;\left({ 1\over A_1} +{1\over A_0} + {m_\pi\over 2 A_1 A_0 m_p}\right)\;.
\label{Tthrnuc}
\end{equation}
Thus $T_{thr,nuc} = 2m_\pi + m_\pi^2/ 2m_p \cong 280$ MeV, as is well
known. This also implies that $T_{thr,nuc} = 171$ MeV/nuc for p-He and $\alpha$-p,
and  $T_{thr,nuc} = 69.9$ MeV/nuc for $\alpha$-He collisions. Concerns \cite{2014PhRvD..90l3014K} about
the accuracy of near-threshold $p + p\rightarrow \pi$ cross sections \cite{1986ApJ...307...47D}
are not as important as  near-threshold contributions of cosmic-ray nuclei.

\section{Concluding remarks}

These few pages should not conceal the much more 
detailed and exhaustive work that must be done
to satisfy cosmic-ray constraints. For this, 
one may consult the detailed GALPROP treatment \cite{2002ApJ...565..280M}, 
which requires processes not often studied,
for example, nuclear energy  losses
from pion production \cite{2015ApJ...802..114K},
or semi-analytic treatments, e.g., \cite{2012JCAP...01..010B}.

The new AMS-02 confirms the hardening in the cosmic-ray
proton and $\alpha$ spectrum reported with PAMELA, and the two data 
sets are typically within $\sim 5$\% of each other.
With this accuracy, and the quality of the Fermi-LAT emissivity data,
renewed focus on metallicity corrections and secondary production 
processes at the $\sim 5$\% level is required.

Before explaining the shape of the LISMCRp spectrum, we must be sure that we
can determine its spectrum from as low an energy, $R\ll 1$ GV ($E_{nuc} \lesssim 100$ MeV/nuc), below
which Solar energetic particles and anomalous cosmic rays start to dominate, 
to as high as energy as possible. It is generally believed that Solar modulation
effects become negligible above $R\sim 10$ GV. From this boundary condition,
we should be able to track the $\gamma$-ray spectrum back to the cosmic rays
and their sources. This is work in progress.


%
%
%

{\bf Acknowledgments}
I thank Drs.\ Elisa Prandini, Simona Toscano, and Andrii Neronov for the kind invitation
to visit Geneva and attend the workshop.  Comments of M.\ Kachelriess, S.\ Ostapchenko,  T.\ Kamae, 
and particularly A.\ Strong are much appreciated. Secondary nuclear production code
used here is available by e-mail request: charles.dermer@nrl.navy.mil.

\end{document}